\documentclass{emulateapj}

\usepackage{amsmath}
\usepackage{xspace} 
\usepackage{graphicx}
\usepackage{txfonts}

\newcommand{\un}[1]{~\hspace{-1pt}\ensuremath{\mathrm{#1}}}
\long\def\symbolfootnote[#1]#2{\begingroup\def\thefootnote{\fnsymbol{footnote}}\footnote[#1]{#2}\endgroup}

\newcommand{\integ}{{\it INTEGRAL}\xspace}
\newcommand{\sgra}{Sgr\,A$^{*}$\xspace}
\newcommand{\am}{$^{\prime}$\xspace}

\newcommand{\aplc}{ApLC\xspace}

\submitted{Received 2003 October 23, accepted 2003 December 1, published 2004 January 23, ApJ 601:L163-L166, 2004 February 1}

\begin{document}

\title{Detection of hard X-ray emission from the Galactic nuclear region with \integ\footnote{Based on observations with INTEGRAL, an ESA project with instruments and science data centre funded by ESA member states (especially the PI countries: Denmark, France, Germany, Italy, Switzerland, Spain), Czech Republic and Poland, and with the participation of Russia and the USA.}}

\author{G. B\'elanger\altaffilmark{1}, A. Goldwurm\altaffilmark{1}, P. Goldoni\altaffilmark{1}, J. Paul\altaffilmark{1}, R. Terrier\altaffilmark{1}, M. Falanga\altaffilmark{1}, P. Ubertini\altaffilmark{2}, A. Bazzano\altaffilmark{2}, M. Del Santo\altaffilmark{2}, C. Winkler\altaffilmark{3}, A.N. Parmar\altaffilmark{3}, E. Kuulkers\altaffilmark{3}, K. Ebisawa\altaffilmark{4}, J.P. Roques\altaffilmark{5}, N. Lund\altaffilmark{6}, F. Melia\altaffilmark{7}}

\altaffiltext{1}{\scriptsize Service d'Astrophysique, DAPNIA/DSM/CEA, 91191 Gif-sur-Yvette, France; \\ 
belanger@cea.fr}
\altaffiltext{2}{\scriptsize IASF CNR, 00133 Roma, Italy; ubertini@rm.iasf.cnr.it}
\altaffiltext{3}{\scriptsize ESA, ESTEC, NL-2200 AG Noordwijk, The Netherlands; \\ 
Christoph.Winkler@rssd.esa.int}
\altaffiltext{4}{\scriptsize ISDC, CH-1290 Versoix, Swizterland; ebisawa@obs.unige.ch} 
\altaffiltext{5}{\scriptsize CESR, Toulouse Cedex~4, France; roques@Sigma-0.cesr.cnes.fr}
\altaffiltext{6}{\scriptsize DSRI, DK-2100 Copenhagen 0, Denmark; nl@dsri.dk}
\altaffiltext{7}{\scriptsize Physics Dept. and Stewart Observatory, Univerity of Arizona, Tucson, AZ 85721, USA; melia@physics.arizona.edu}

\begin{abstract}
This Letter presents the first results of an 
observational campaign to study the Galactic center with \integ,
the {\it Internaltional Gamma-Ray Astrophysics Laboratory}.
Mosaicked images were constructed using data obtained with
ISGRI, the soft gamma-ray instrument of the coded aperture IBIS imager,
in the energy ranges 20--40 and 40--100\un{keV}.
These give a yet unseen view of the high-energy sources
of this region in hard X-rays and gamma-rays with 
an angular resolution of 12\am (FWHM).
We report on the discovery of a source, IGR J1745.6--2901, 
coincident with the Galactic nucleus \sgra to within 0\hspace{1pt}\am\hspace{-5pt}.9. 
Located at 
R.A.(J2000.0) = $\mathrm{17^{h}45^{m}38^{s}\hspace{-2pt}.5}$, 
decl.(J2000.0) = $-29^{\circ}01'15''$,
the source is visible up to about 100\un{keV} with a 
20--100\un{keV} luminosity at 8\un{kpc} of 
$(2.89 \pm 0.41) \times 10^{35}$\un{ergs\; s^{-1}}.
Although the new \integ source 
cannot unequivocally be associated to the Galactic nucleus, 
this is the first report of significant hard X-ray emission from within 
the inner 10\am of the Galaxy and
a contribution from the Galactic supermassive black hole 
itself cannot be excluded.

\end{abstract}


\section{Introduction}

From the prediction of the existence of a massive,
compact source at the center of the Milky Way 
by Lynden-Bell \& Rees (1971) 
to the discovery of such a source in the radio domain 3 years later by 
Balick \& Brown (1974), 
and to the first detection of soft X-rays 
unmistakably attributable to it by 
{\it Chandra} in 1999 \citep{c:baganoff03a},
the supermassive black hole candidate \sgra 
and the Galactic center (GC) region as a whole
have been put under intense scrutiny for many 
years from radio wavelengths to gamma-rays.
This has led to several discoveries, and 
advances in our understanding of the processes
and interactions at the heart of the Milky Way.
For example, it is now known that the soft X-ray emission
($<$10\un{keV}) in the central 10\am is heavily dominated by diffuse 
emission due primarily to hot gas \citep{c:koyama96,c:sidoli-mereghetti99},
and that only about 10\% 
of the total emission in this energy range
can be accounted for by X-ray point sources brighter than 
$10^{31}\un{ergs\; s^{-1}}$ \citep{c:muno03a}.
Also, the contribution of point sources in this domain is about the same
along the Galactic ridge as it is in the GC \citep{c:ebisawa01}.

The Galactic nuclear region consists of six primary components that give
rise to an array of complex phenomena
through their mutual interactions.
These constituents are: \sgra,
a supermassive black hole with a mass of 
around $3.5 \times 10^6 \un{M_{\odot}}$ \citep{c:schodel02,c:ghez03a}; 
the surrounding cluster of evolved and young stars;
ionized gas streamers, 
some of which form a three armed spiral
centered on \sgra known as Sgr\,A\,West;
a molecular dusty ring surrounding Sgr\,A~West;
diffuse hot gas; and a powerful supernova-like remnant 
known as Sgr\,A~East \citep{c:melia01}.
Furthermore, 
both the IR and the X-ray point source populations
decreases in spatial density approximately as $1/R^{2}$, 
where $R$ is the distance from the GC \citep{c:serabyn96,c:muno03a}.

The first imaging observations of the GC 
in hard X-rays were performed by the X-ray telescope on
{Spacelab\,2} in the range 2.5--20\un{keV} with an angular resolution of 3$'$.
A source located within 
1\am\hspace{-6pt}.1 of \sgra was detected \citep{c:skinner87}.
Such a source was also detected by ART-P (8--20\un{keV}) 
on the {\it Granat} satellite in 1990-1991 \citep{c:pavlinsky94}. 
At energies above 20\un{keV},
only the coded mask instrument SIGMA/{\it Granat},
sensitive to energies above 35\un{keV},
provided imaging capability with a $\sim$20\am angular resolution,
allowing an exploration of the dense GC region.
In spite of the deep $9 \times 10^{6}\un{s}$ SIGMA survey of the
central parts of the galaxy performed between 1990 and 1997, 
only upper limits were set for 
the hard X- and gamma-ray emission from the neighborhood of \sgra
at energies above 35\un{keV} \citep{c:goldwurm94,c:goldoni99}.
The derived low bolometric luminosity of the Galactic nucleus (GN),
in contrast with the powerful output from active galactic nuclei
or black hole binaries,
has motivated the developement of several models for
radiatively inefficient accretion onto or ejection from \sgra.
These models have been widely applied 
to other accreting systems \citep{c:melia01}.

A recent breakthrough discovery by
{\it Chandra} and {\it XMM-Newton} is that \sgra is the source of 
powerful X-ray flares \citep{c:baganoff01,c:goldwurm03a,c:porquet03}
during which the soft X-ray luminosity can increase 
by factors of 50--180 over a period of up to 3 hr.
Some of these X-ray flares feature a significant hardening of the spectrum 
up to photon indices of $\sim$1.
Furthermore, a very recent discovery with 
the Very Large Telescope NAOS/CONICA imager \citep{c:genzel03}
and the Keck telescope \citep{c:ghez03b}, 
is that \sgra is also the source of frequent IR flares.
This activity could indicate the presence of
an important population of non-thermal electrons in the 
vicinity of the black hole. 
These results have raised great interest in the possibility
of observing hard X-rays from the GN,
a measure of which may particularly shed light 
on the relative role of accretion and ejection 
in the \sgra system.

This Letter presents preliminary results of an observation campaign
to study the GC at high energies with \integ, the  
{\it Internaltional Gamma-Ray Astrophysics Laboratory},
performed during the 
\integ Galactic Center Deep Exposure (GCDE) program 
in the spring of 2003.
The focus is set on the inner 10\am--15\am of the Galaxy
and, more particularly, 
on the detection of a significant excess in this region.
The results are based on data obtained with the \integ
Soft Gamma-Ray Imager (ISGRI), the low-energy  
camera of the main Imager on Board the \integ Satellite (IBIS),
proven to be accurate between 20 and 200\un{keV}.
The morphology of the central 2$^{\circ}$ in the 20--100\un{keV} band
and the derived positions of closeby high-energy sources are 
briefly discussed to elucidate the experimental context from which
the results on the central source are drawn. 

\begin{figure}[htb]
    \epsscale{1.1}
\plotone{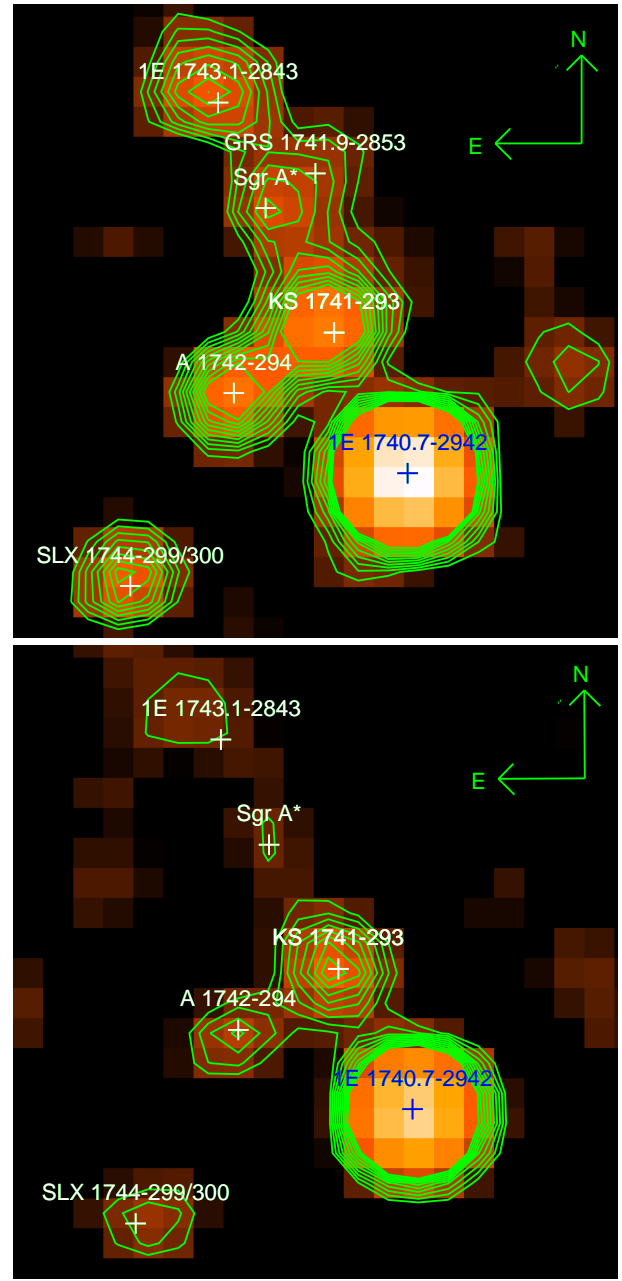}
\caption{\footnotesize Milky Way's center seen in a 
	$2^{\circ}\hspace{-2pt}\times 2^{\circ}$ field
	by the IBIS/ISGRI instrument
 	in the energy ranges 20--40\un{keV} (top)
	and 40--100\un{keV} (bottom).
        Each image pixel size is equivalent to about 5\un{arcmin}.
	Ten contour levels mark iso-significance
	linearly from about 4$\sigma$ to 15$\sigma$.
        \label{f:sgra}}
\end{figure}

\section{Observations and Results}

\integ \citep{c:winkler03} 
is a European Space Agency observatory that
began its mission on 2002 October 17 carrying four instruments.
These consist of two main ones, 
IBIS \citep{c:ubertini03} and SPI, 
the Spectrometer on \integ \citep{c:vedrenne03};
and two monitors, JEM-X \citep{c:lund03} and OMC \citep{c:mas-hesse03}.

The IBIS coded mask instrument 
is characterized by a wide field of view (FOV) of
$29^{\circ} \hspace{-1pt} \times 29^{\circ}$, 
($9^{\circ} \hspace{-1pt} \times 9^{\circ}$ fully coded),
a point spread function (PSF) of 12\am (FWHM),
and a sensitivity over the energy range between 15\un{keV} and 8\un{MeV}.
This sensitivity is achieved via two detector layers: 
ISGRI \citep{c:lebrun03}, an upper CdTe layer 
sensitive between 15\un{keV} and 1\un{MeV} 
with peak sensitivity between 15 and 200\un{keV}, 
and PICsIT 
a bottom CsI layer, sensitive between 200\un{keV} and 8\un{MeV}.

We have analyzed IBIS/ISGRI data 
collected between 2003 February 28 and May 1
in a series of fixed pointings lasting about 37 minutes each.
These include all GCDE data for which the pointings
include the GN ($\sim\! 625\un{ks}$), 
and two Target of Opportunity observations ($\sim\! 475\un{ks}$). 
Data reduction was performed using the standard OSA\,2
\integ Science Data Center analysis software \citep{c:goldwurm03b},
whose present version of the analysis procedures and calibration
files do not allow for a full correction of systematic effects.
Thus, in order not to over estimate the detection level of a source,
taking into account fluctuations in the observation-dependent 
background noise levels as well as systematics,
the significance was normalized to the fitted width of the distribution of 
individual significances in the image.
This straight forward normalization procedure ensures that 68\%
of the distribution of significance values in the image are
indeed contained within 1$\sigma$ of the mean.
Image reconstruction can be summarized as follows:
from the events list for a pointing, subsets of
events are selected according to energy bins.
Each subset is used to build a detector image or shadowgram. 
Convolution of the shadowgram with a decoding array
gives rise to a sky image containing the main peak 
of all sources in the FOV and their secondary lobes. 
Source identification and subtraction of secondary lobes
results in the final reconstructed sky image.
Fluxes are derived using \integ observations
of the Crab Nebula performed just prior to the start of the GCDE,
and luminosity calculations are based on
a distance of 8\un{kpc} to the GC 
assuming a Crab spectrum of a power law
with photon index 2.12 and flux density
at 1\un{keV} of 8\un{ph\:cm^{-2}\:s^{-1}\:keV^{-1}} \citep{c:bartlett94}.

The maps of the GC shown in Figure~\ref{f:sgra} 
were constructed by summing the reconstructed images of 571 individual
exposures. 
The total effective exposure time is
about $8.5\times 10^{5}\un{s}$ on the GN. 
In these signal-significance maps of the central two degrees
of the Galaxy where ten contour levels mark iso-significance 
linearly from about $4\sigma$ up to $15\sigma$, 
we can see what appear to be six distinct sources:
1E~1740.9--2942.7, KS~1741--293, A~1742--294,
1E~1743.1--2843, SLX~1744--299/300, whose nominal positions
are marked by crosses and identification is still preliminary, 
and a source coincident with the radio position of \sgra.
1E~1740.7--2942 is a black hole candidate,
KS~1741--293 and A~1742--294 are 
neutron star Low-Mass X-Ray Binary (LMBX) bursters,
SLX~1744--299/300 are in fact two LMXBs separated 
by only 2.72\am, 
and 1E~1743.1--2843 is an X-ray source
whose nature is still uncertain.
In the 20--40\un{keV} band, contours of the central source
peak at the position of \sgra with a significance level of $8.7\sigma$ 
but are elongated toward GRS~1741.9--2853.
This suggests some contribution from this 
transient neutron star LMXB burster system 
recently observed to have returned to an active state \citep{c:muno03b}
but could also be due to an uncorrected background structure.
The central source is also marginally visible in the 40--100\un{keV} band 
at a level of $4.7\sigma$ but without any contribution from the direction of 
GRS~1741.9--2853.

\begin{deluxetable}{lcccc}
\tablecolumns{5} 
\tablewidth{0pc} 
\tablecaption{\integ source position location accuracy} 
\tablehead{ 
\colhead{Source name} &  \colhead{Significance} & \colhead{Fitted position}   & \colhead{Offset$^{~\mathrm{a}}$} \\
		      & \colhead{}& \colhead{(R.A., decl.)}  &  \colhead{(arcmin)} }
\startdata
1E 1740.7-2942		&  70.0  		& $265.98, -29.74$ 		& 0.32 \\
KS 1741-293		&  18.9  		& $266.23, -29.32$ 		& 2.31 \\
A 1742-294		&  15		 	& $266.52, -29.51$ 		& 0.05 \\
SLX 1744-299/300	&  9.6	 		& $266.86, -30.02$ 		& 1.59 \\
1E 1743.1-2843		&  9.2	 		& $266.59, -28.67$ 		& 3.96 \\
\sgra			&  8.7 	 		& $266.41, -29.02$ 		& 0.86 \\
\enddata 
\tablenotetext{a}{Distance between the fitted and nominal source positions} 
\label{t:sourcePositions}
\end{deluxetable} 

The position and flux of the central excess
in the 20--40\un{keV} map were determined
by fitting the peaks with a function 
approximating the instrument's PSF \citep{c:gros03}
in two different ways:
(1) all the emission is attributed to one source
and is fitted as such to determine its peak height and position, 
(2) the emission is attributed to two sources:
a new source and GRS~1741.9--2853, whose position is then fixed
to the one determined with the {\it Chandra} observatory \citep{c:muno03b}. 
Both of these involve a simultaneous fit of all the sources 
in the 2 deg$^{2}$ field of the GC.
In the first case, 
we obtain a source position of 
R.A.(J2000.0)=$\mathrm{17^{h}45^{m}22^{s}\hspace{-3pt}.5}$,  
decl.(J2000.0)=$-28^{\circ}58'17''$,
and a flux of 
about 5.4 mcrab or $(3.21 \pm 0.36) \times 10^{-11}\un{ergs\; cm^{-2}\; s^{-1}}$.
In the second,
the position is 
$\mathrm{17^{h}45^{m}38^{s}\hspace{-3pt}.5}$,  
$-29^{\circ}01'15''$,
and the flux is  
about 3.2 mcrab or 
$(1.92 \pm 0.36) \times 10^{-11}\un{ergs\; cm^{-2}\; s^{-1}}$.
The central source's 40--100\un{keV} peak 
position is in very good agreement with the one determined 
using the second method outlined above, and since there is clearly no 
visible contribution from a neighboring source,
the 40--100\un{keV} flux was extracted at that position.
The estimated error on the position is of about 4\am 
for a detection at the significance level of 8.7$\sigma$.
These positions are respectively 4\am\hspace{-6pt}.6 and 
0\am\hspace{-5pt}.9 from the radio position of \sgra.
The results of the fine position determination for
the six above mentioned GC sources are shown in
Table~\ref{t:sourcePositions}. 
In Table~\ref{t:fluxes}, flux estimates 
for the central source are given
adopting the result of the second method for 
position and flux determination.
The hardness ratio (HR) --- ratio of the count rate in the
high-energy band over that in the low-energy band ---
for the detected excess is $0.90 \pm 0.20$. 
As a possible indication of the nature of the detected excess,
we can compare the values of the HR to the two brightest
sources in the field.
The BH candidate 1E~1740.7--2942 has a HR of $1.20 \pm 0.03$,
and the neutron star LMXB KS~1741--293 has a HR of $0.89 \pm 0.08$.

\begin{figure}
      \epsscale{1.0}
\plotone{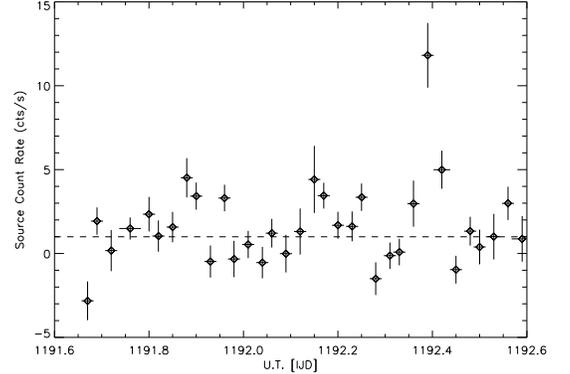}
      \caption{\footnotesize Portion of the total 20--40\un{keV}
	lightcurve obtained by extracting the
	flux at the position of \sgra in the background corrected 
	intensity map of each elementary exposure. The statistical 
	errors are shown, and the dashed line corresponds to the
	mean count rate over the entire data set.
	\label{f:lightCurve}}
\end{figure}

The excess appears to be somewhat
variable throughout the observations. 
In particular, we have detected a sudden increase in 
flux where the count rate rose by a factor of $\sim\! 12$ 
with respect to the mean for about 40\un{min}.
Figure~\ref{f:lightCurve} shows a portion of the light curve from
the estimated 20--40 keV flux at the radio position of \sgra. 
Each data point corresponds to the average flux and 
statistical error from $\sim$37 minute intervals during observations that lasted 
 $\sim$19 hr on 2003 April 6.
Universal time is reported in \integ Julian days, 
i.e. days from 2000 January 1.  
The dashed line depicts the mean count rate  over
the entire data set. 
One peak clearly stands out at 
a pure statistical significance of 5$\sigma$, 
bearing in mind that this significance  may be hampered by systematic effects.
A simultaneous increase in flux is seen with a lower significance
in the 40--100 keV light curve.
In what follows we discuss a number of possible
explanations for this detected excess.

\section{Discussion and Conclusion}

\begin{deluxetable}{llcc}
\tablecolumns{4} 
\tablewidth{0pc} 
\tablecaption{Measured and extrapolated fluxes} 
\tablehead{ 
\colhead{Instrument} & \colhead{Energy}   & \multicolumn{2}{c}{Energy Flux} \\
\colhead{} & \colhead{(keV)}	& \colhead{($10^{-11}\un{ergs\; cm^{-2}\; s^{-1}}$)} & \un{(mCrab)} 	
}
\startdata 
\integ		& 20--40$^{\mathrm{a}}$	& 	$1.92 \pm 0.36$		& $3.2 \pm 0.6$ \\
(IGR 1745.6--2901)& 40--100$^{\mathrm{a}}$&	$1.86 \pm 0.40$		& $3.4 \pm 0.7$ \\
\hline
		& 2--10$^{\mathrm{a}}$	&		10.0		& 5.78		\\
{\it XMM-Newton}& 20--40$^{\mathrm{b}}$	&		0.61	   	& 1.03		\\
		& 40--100$^{\mathrm{b}}$&		0.031	   	& 0.04		\\
\hline
		& 2--10$^{\mathrm{a}}$	&		11.0	   	& 6.36		\\
{\it BeppoSAX}	& 20--40$^{\mathrm{c}}$	&		0.89	   	& 1.49		\\
		& 40--100$^{\mathrm{c}}$&		0.061		& 0.09
\enddata
\tablenotetext{a}{Measured}
\tablenotetext{b}{Extrapolated using kT = 7.2 keV}
\tablenotetext{c}{Extrapolated using kT = 8.0 keV}
\label{t:fluxes}
\end{deluxetable}

In attempting to identify the excess at
$\mathrm{17^{h}45^{m}38^{s}\hspace{-2pt}.5}$, 
$-29^{\circ}01'15''$,
we first 
consider which sources of high-energy emission 
can be excluded based solely on ISGRI's error radius of 4\am.
There are four such sources:
the radio arc located about 15\am from the GN;
the non thermal X-ray filaments associated with it 
and the dense molecular cloud {G0.13--0.13} \citep{c:yusef02};
GRS 1743--290, observed by SIGMA in 1991 \citep{c:goldwurm94};
and GRS 1741.9-2853 mentioned above and clearly unable to account 
for the bulk of the excess.

Second, 
we argue that this detection cannot be attributed to 
both diffuse emission and point sources as thus far observed
at low X-ray energies within 8\am--10\am of \sgra.
This is so because it is incompatible with 
the extrapolation at higher energies of 
the flux integrated over this region 
as measured by {\it XMM-Newton} (A. Decourchelle 2003, private communication) 
and {\it BeppoSAX} \citep{c:sidoli99}.
The analysis of 2001 {\it XMM-Newton} data in the 2--10\un{keV} range
consisted of integrating the total X-ray flux over a
radius of 10\am around \sgra.
The resulting spectrum is best fitted by 
a two-temperature plasma model 
with $\mathrm{kT_{1}}$=0.6\un{keV} for the soft component
and $\mathrm{kT_{2}}$=7.2\un{keV} for the harder one.
1997 {\it BeppoSAX} observations of the Sgr\,A complex over
a radius of 8\am around \sgra in the same energy band 
agree with {\it XMM-Newton}'s results.
In this case
the data are also best described by a two-temperature
model with $\mathrm{kT_{1}}$=0.6\un{keV} and
$\mathrm{kT_{2}}$=8\un{keV}.
The extrapolation of the {\it XMM-Newton} and {\it BeppoSAX} results
to higher energies was done using kT=7.2 and 8\un{keV},
respectively.
Both instruments' direct measurements
of the diffuse X-ray flux
are consistent with those of {\it Chandra} \citep{c:baganoff03a} 
and ART-P \citep{c:sunyaev93}.
Table~\ref{t:fluxes} reports the {\it XMM-Newton} and {\it BeppoSAX} 
measurements as well as their respective extrapolations.
Note that the variability timescale of the \integ source 
equally excludes a major diffuse contribution 
and also strongly suggests that
accelerated charged particles in the expanding
shell of Sgr\,A East \citep{c:melia98} cannot 
significantly contribute to the detected signal.

Third, 
since the detected excess
cannot be fully accounted for by diffuse emission
and is variable,
it is reasonable to consider it a point source, IGR~J1745.6--2901, 
of 20--100\un{keV} luminosity $\sim$$3 \times 10^{35}\un{ergs\; s^{-1}}$.
One known point source found within 4\am of IGR~J1745.6--2901
that could contribute to the signal is
the eclipsing burster 
AX J1745.6--2901 \citep{c:maeda96,c:sakano02}.
This source lies about 1\am.3 from \sgra and 
was detected in a high state only once, in 1994 \citep{c:sakano02}.
The extrapolation of the flux measured in that 
state is marginally compatible with that of IGR J1745.6--2901,
and this only in the low-energy range.
The last confirmed detection of AX J1745.6--2901 was in 1997 by ASCA,
when it was found in a much lower flux state. 
It may have recently been detected by {\it Chandra}.
A timing analysis of the ASCA data on AX J1745.6--2901
revealed a modulation with a 40\% decrease in flux 
and period of $8.356 \pm 0.008 \un{h}$ \citep{c:maeda96}.
1997 ASCA data folded with that period show a slight
dip, but since phase information could not be preserved over
the 2.5 yr that separated the two observations,
no unique solution for the period was found \citep{c:sakano02}.
Folding the \integ data for IGR~J1745.6--2901 with a period 
of 8.356 hr does not reveal any modulation.

{\it Chandra} observations of the GC performed on 
2003 June 19 indicate that two sources,
CXOGC J174540.0--290005 and CXOGC J174540.0--290014, 
found within 0\am\hspace{-6pt}.4 of \sgra, 
and another, CXOGC J174535.5--290124, lying 1\am\hspace{-6pt}.3 from \sgra 
and possibly the same as AX~J1745.6--2901,
had a combined intensity in the 2--8\un{keV}
band that was more than 30 times that of \sgra in its quiescent
state on that day (F.~K. Baganoff 2003, private communication).
The GN was not particularly bright on June 19,
but it must have been somewhat more active in May
when an intense IR flare was observed to come from a region 
possibly very near the supermassive black hole's 
event horizon \citep{c:genzel03}.

It is noteworthy that the fairly hard X-ray source that ART-P detected 
in 1990 at the position of \sgra, had a 8--20\un{keV} flux 
that varied between about 5 and 11\un{mcrab} \citep{c:pavlinsky94}.
This is consistent with our detection in the 20--40\un{keV} energy range.
We also recall the 2~$\sigma$ upper limits set by SIGMA
on the 40--80\un{keV} and 80--150\un{keV} luminosity of the GN: 
$3.4\times 10^{35}$ and
$2.8\times 10^{35}\un{ergs\; s^{-1}}$ respectively \citep{c:goldoni99},
both of which are compatible with our detection.

Finally, \integ has detected a very hard X-ray source 
at a position coincident to within 1\am with \sgra.
Because of IBIS's 12\am angular resolution,
the emission from IGR~J1745.6--2901 cannot be  
attributed to one specific object and could be 
made up of several contributing sources 
found within a couple of arcminutes of \sgra.
Nonetheless, this remains the first detection of
emission at energies greater than 20\un{keV} 
from the very close vicinity of the GN,
and a contribution from \sgra itself cannot be excluded. 

More INTEGRAL data are needed to better constrain the 
position, spectral shape, variability properties and the possibly
multiple nature of IGR~J1745.6--2901. 
We expect that the most constraining results will be provided by 
simultaneous observations in hard and soft X-rays 
with INTEGRAL and {\it XMM-Newton} or {\it Chandra}.

\acknowledgements{
We gratefully thank Fran\c{c}ois Lebrun for insightful suggestions
and Anne Decourchelle for her analysis of {\it XMM-Newton} data on the GN.

\end{document}